\documentclass[floatfix,twocolumn,showpacs,prb]{revtex4}
\usepackage{graphicx}
\usepackage{amsmath}
\usepackage{bm}
\usepackage{dcolumn}

\begin{document}

\title{Competing magnetic anisotropies in atomic-scale junctions}

\author{Alexander~Thiess$^{1}$}
\author{Yuriy~Mokrousov$^{1}$}
\email{y.mokrousov@fz-juelich.de}
\author{Stefan~Heinze$^{2}$}
\affiliation{$^1$Institut f\"ur Festk\"orperforschung,
Institute for Advanced Simulation, Forschungszentrum J\"ulich,
D-52425 J\"ulich, Germany}
\affiliation{$^2$Institute of Theoretical Physics and Astrophysics,
Christian-Albrechts-University of Kiel, D-24098 Kiel, Germany}
\date{\today}
\begin{abstract}
Using first-principles calculations, we study the magnetism of
$5d$ transition-metal atomic junctions including structural
relaxations and spin-orbit coupling. Upon stretching monatomic
chains of W, Ir, and Pt suspended between two leads, we find
the development of strong magnetism and large values of the
magnetocrystalline anisotropy energy (MAE) of up to 30~meV~per 
chain atom. We predict that switches of the easy magnetization axis 
of the nanocontacts upon elongation should be observable by 
ballistic anisotropic magnetoresistance measurements. Due to the 
different local symmetry, the contributions to the MAE of the 
central chain atoms and chain atoms in the vicinity of the leads 
can have opposite signs which reduces the total MAE. 
We demonstrate that this effect occurs independent of the
chain length or geometry of the electrodes.
\end{abstract}


\maketitle

\section{Introduction}

Fascinating insights into the formation of atomic chains
consisting only of a few atoms suspended between two electrodes
have been obtained by transmission electron microscopy and
mechanically controllable break junction techniques.\cite{Sokolov,
Smit:01.1,Kizuka} These experiments triggered the imagination
of scientists to use such atomic-scale junctions as future
electronic devices by exploiting their unique properties,
e.g.~ballistic electronic transport.\cite{Sokolov,Velev:05.1,
Smogunov:08.1} Due to their reduced dimensions, atomic chains
have also been predicted to develop magnetic moments even for
elements nonmagnetic in bulk such as Ir, Pt, or
Pd.\cite{Delin:03.1,Delin:04.1} Recently, an indirect proof 
has been given that transition-metal chains in break junctions 
are in general magnetic.\cite{m-supp} Mastering the 
intriguing magnetic properties of suspended
chains would enable spintronic applications based on the
unique possibility to probe, control and switch the magnetic
states by spin-polarized currents.\cite{Lucignano,Calvo}

The orientation of the magnetic moments is stabilized against
thermal fluctuations by the magnetocrystalline anisotropy
energy (MAE) arising from spin-orbit coupling. Theoretical
studies of this key quantity performed for idealized systems
such as free-standing infinite monowires (MWs) and small clusters
suggest giant values for $4d$- and $5d$-transition-metals
and switching of the easy axis upon stretching the wires.
\cite{Mokrousov:06.3,MacDonald1,Ferrer:07.1} Even the size
of the magnetic moment itself can crucially depend on the
magnetization direction, an effect coined as colossal magnetic
anisotropy.\cite{Smogunov:08.1} The ballistic conductance in
such wires varies with the orientation of the magnetization
direction with respect to the chain axis giving rise to
ballistic anisotropic magnetoresistance (BAMR).\cite{Velev:05.1,
Sokolov} While the predicted effects probably occur in real 
atomic-scale junctions, the theoretical studies have either 
focused on idealized systems or neglected varying interatomic 
distances in the chains.\cite{Smogunov:08.2,Smelova:08.1,Czerner:08.1}
However, for small suspended chains variations in the interatomic
distance due to stretching the electrodes are unavoidable and
can provide a unique way to control the properties of the system,
not available in higher dimensions.\cite{Thiess:08.1,Smogunov:08.1,
Delin:03.1,Mokrousov:06.3}

Here, we go beyond such theoretical studies by explicitly
taking the contacts into account and by investigating the
development of magnetism in atomic-scale junctions upon
increasing the distance between the electrodes. We perform
first-principles calculations for W, Ir and Pt junctions
including spin-orbit coupling and structural relaxations
and focus on atomic chains of three atoms suspended between
two bulk-like bcc-(001) electrodes as shown in Fig.~\ref{fig1}.
These transition-metals are common tip materials in scanning
tunneling microscopy, and Ir and Pt have been shown to form
long atomic chains in break junction experiments.\cite{Smit:01.1}
In particular we concentrate on the influence of varying
tip-to-tip separation on the magnetic properties of the
junctions. We observe
that relaxations of the chain atoms upon stretching are in
accordance to observed expermentally trends of chain formation
for these elements.\cite{Thiess:08.1} We also find
that large spin moments develop in the suspended trimers upon
stretching and on the central atom they reach similar values
as in infinite monowires.\cite{Delin:03.1,Delin:04.1,m-supp} 
Interestingly, the W trimer behaves as a single spin impurity 
as only the central atom develops a sizable magnetic moment.

By including spin-orbit coupling (SOC), we obtain the MAE of
the suspended chains and observe large values of 10$-$30 meV 
per atom. Surprisingly, we find that the contribution to
the MAE from central chain atoms and atoms in contact with
the electrodes can have opposite signs and are of similar
giant magnitude of a few tens meV per atom.
The magnitude of this effect, unknown in bulk or on surfaces 
may lead to a strong tendency towards non-collinear magnetic 
order in atomic-scale junctions even when the exchange 
interactions are small or profoundly collinear.\cite{Lichten} 
We prove the generality of this phenomenon by showing that
it occurs for chains of different length with different geometry
of the contacts. Further, the MAE may exhibit changes of sign upon 
stretching,~i.e.~the easy magnetization direction switches 
from along the chain to perpendicular to it. We demonstrate 
that switching may be observable experimentally by 
measuring the ballistic conductance of the junctions. 

\section{Method and Structure}

\begin{figure}[t!]
\begin{center}
\includegraphics[scale=0.39]{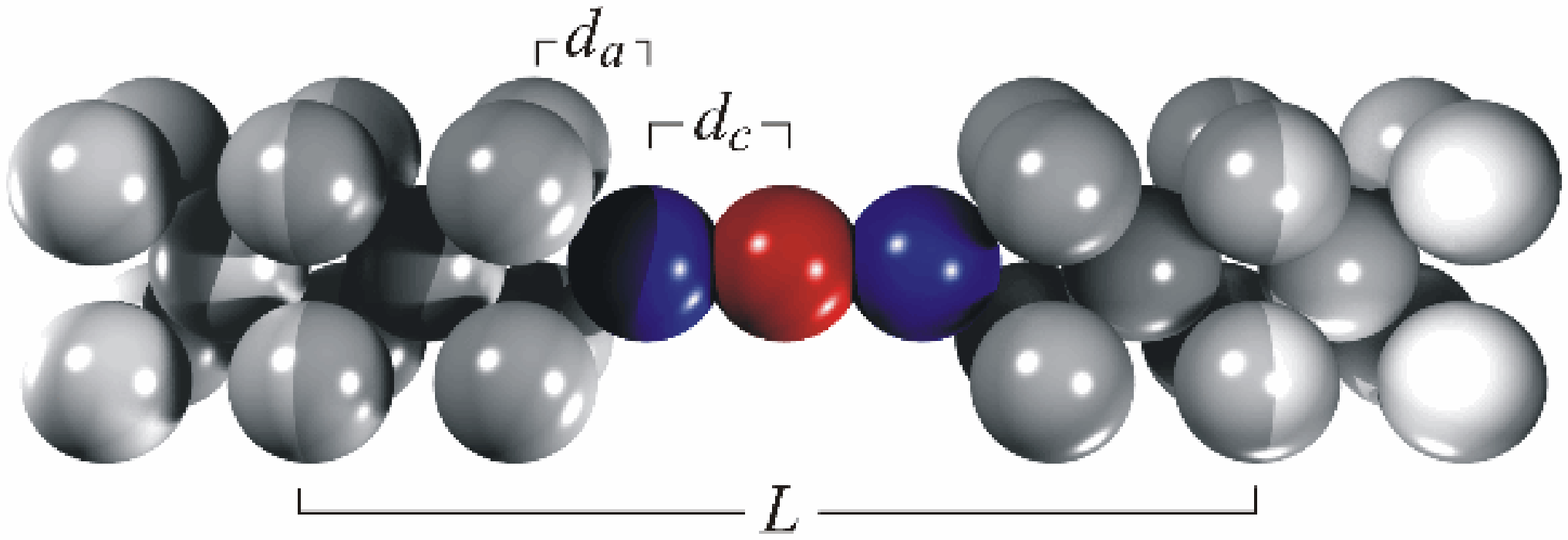}\\[0.5cm]
\includegraphics[scale=0.35]{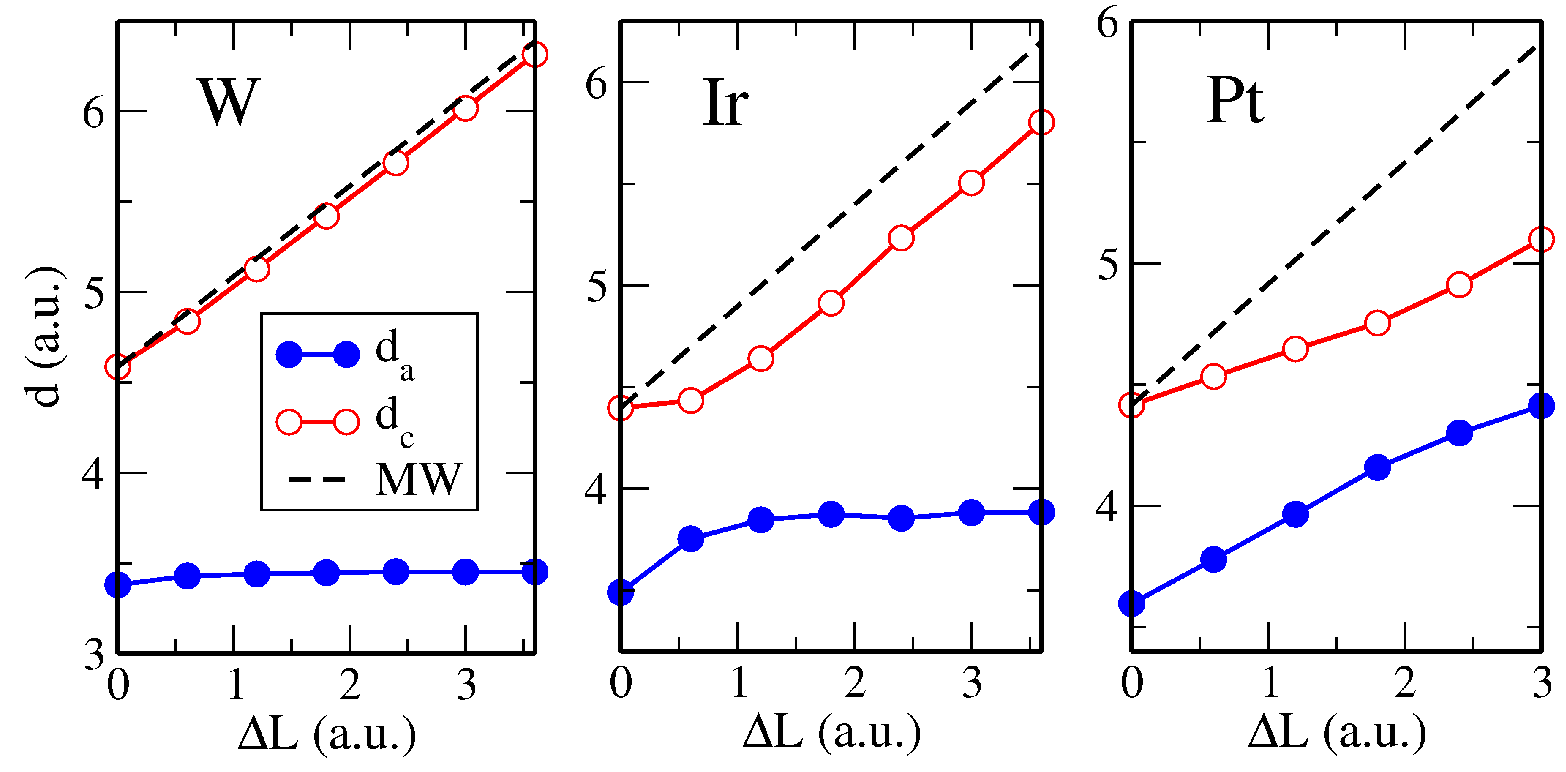}
\end{center}
\caption{\label{fig1} (color online)
Upper panel: three-atoms chain suspended between two
bcc(001)-contacts. $L$ denotes the length of the unit cell
consisting of 17 atoms, while $d_a$ and $d_c$ give the distance
between the first plane of the contact to the apex atom and
from the apex atom to the central atom of the trimer,
respectively. Lower panel: relaxation of the $d_c$ and $d_a$
bonds in suspended trimers of W, Ir, and Pt upon increasing
$L$. For comparison, the bond length for ideal stretching
with fixed apex atom positions is given (dashed line).}
\end{figure}

We have performed first-principles calculations within the
generalized gradient approximation (GGA) to the density
functional theory (DFT). We employed the full-potential linearized
augmented plane-wave method for one-dimensional (1D) systems
(1D-FLAPW),\cite{Mokrousov:05.1} as implemented in the
\texttt{FLEUR} code.\cite{fleur} Basis functions were
expanded up to $k_{\rm max} = 4.0$~a.u.$^{-1}$ and we have
used 8 $k$-points in one half of the 1D Brillouin zone.
Considering more $k$-points does not change the values of
the spin moments and magnetic anisotropy energies significantly.
The contact structure has first been relaxed separately in
all three dimensions and then kept fixed for all contact
separations $L$ (Fig.~1). The apex atoms of the
trimer have been relaxed along the chain direction ($z$)
upon varying $L$, while the central atom is
fixed by symmetry (Fig.~\ref{fig1}). Calculations have been
performed for junctions stretched from the equilibrium up to
$\Delta L$ = 3~a.u. For Pt and Ir trimers we considered the 
ferromagnetic ground state, while W trimer always converges
to the antiferromagnetic solution.

Spin-orbit coupling SOC was included into the calculations
in a non-perturbative way.\cite{Freeman} 
We define the magnetic anisotropy
energy (MAE) as the total energy difference between
configurations with a magnetization along the chain axis
($z$-direction), and perpendicular to it ($r$-direction).
The MAE was calculated employing the force theorem, and
for several electrode stretchings, $\Delta L$, we checked
the force theorem results for the trimer's MAE against 
self-consistent calculations, finding good qualitative 
agreement. Taking into account the magnitude of the 
obtained MAE we can safely neglect the magnetic 
dipole-dipole contribution.

In order to check the applicability of our 1D geometry for the 
electrodes, which allows us to efficiently perform the 
demanding calculations with structural relaxations and 
including SOC, we have also considered a geometry with
electrodes represented by surfaces, c.f.~Fig.~2(a). Due 
to the immense computational cost, we have restricted our 
test to the most critical case of a Pt trimer and chose an 
elongation of $\Delta L=2.4$~a.u. For these calculations 
we used the film version of the \texttt{FLEUR} code. We
performed a single super-cell calculation in which the trimer
was suspended between two electrodes consisting of a 
four Pt atom square base deposited onto three layers of fcc Pt(001).
This configuration will be referred to as "2D" in the following, Fig.~2.
The ideal bulk values of Pt with an in-plane lattice constant of 
5.23~a.u.~were taken for the four contact atoms and the bulk-like 
leads. The positions of the trimer's atoms were set to their 
relaxed values at $\Delta L=2.4$~a.u.~obtained within the 
wire geometry of Fig.~1. The in-plane separation between
the neighboring trimers was set to 15.7~a.u. We used a
basis functions cut-off of $k_{\rm max}=3.7$~a.u.$^{-1}$ and 
15 $k$-points in the irreducible part of the 2D Brillouin zone,
while 144 $k$-points in the whole Brillouin zone were used
in order to determine the MAE with the force theorem.
Along the trimer's axis only the $\Gamma$-point was used.

\begin{figure}[ht!]
\begin{center}
\includegraphics[scale=0.35]{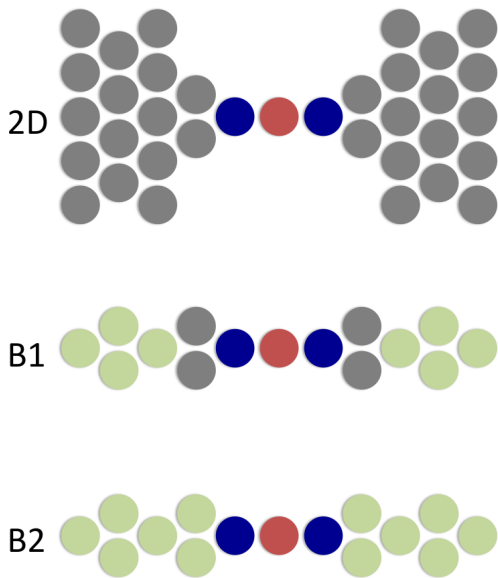}
\end{center}
\caption{\label{fig22} (color online)
2D configuration, and B1 and B2 cases of the Pt trimer 
considered for the separation of Pt electrodes of $\Delta L=2.4$ 
and 1.8~a.u., respectively. In 2D configuration the trimer is 
suspended between two electrodes consisting of a four Pt 
atom square base deposited onto three layers of fcc Pt(001). Green 
color in B1 and B2 marks the Pt atoms of the contacts in which 
the spin moment was artificially quenched. In accordance to Fig.~1, 
red color marks the central atom and the blue color marks the apex 
atoms of suspended trimer. For more details see text.}
\end{figure}

\section{Relaxations and Magnetism}

Under the condition of reduced coordination of transition metal
atoms in a monoatomic chain suspended between thicker leads, 
a situation realized~e.g.~in a break junction experiment, the 
distance $d$ between the chain atoms becomes of utter 
importance. Upon pulling apart the leads during the chain 
elongation process\cite{Kizuka,Thiess:08.1} $d$ is constantly 
changing which has dramatic consequences for the magnetic 
properties of suspended chains, in particular for heavy $4d$ 
and $5d$ transition-metals.\cite{Smogunov:08.1,Delin:03.1,
Delin:04.1,Mokrousov:06.3} By increasing the separation 
$\Delta L$ between the leads and accounting at the same 
time for the changes in the chain's interatomic distances 
$-$ which was neglected in most of the previous {\it ab initio} 
studies of magnetism in suspended chains $-$ we aim at 
mimicking a real break junction experiment.

The relaxations of the trimers along the chain's symmetry 
$z$-axis are shown in Fig.~\ref{fig1}, where the bond lengths 
$d_c$ and $d_a$ are presented as a function of contact 
separation $L$. For W with an almost half-filled $5d$-shell
and large bulk cohesive energy almost the entire increase in
contact separation enters the $d_c$ bond length, while the
$d_a$ bond between the apex atom and the contact is nearly
unaffected. In contrast, for Ir and Pt, some of the stretching
transfers into an increased bond length at the contact.
The apex atom of the Ir trimer has a stronger bond to the
contact than to the central atom, which is reflected in the
linear regime of $d_c(L)$ for $\Delta L > 1.5$~a.u.
For Pt, the $d_c$ bond is significantly stronger resulting in 
a remarkable stretching of the $d_a$ bond even at large 
electrode separations, which in a realistic break junction will 
eventually result in chain elongation.\cite{Thiess:08.1} Overall, 
our predicted trend is in accordance with experimental 
observations of increased probability for chain formation 
when going from W to Pt.\cite{Smit:01.1,Untiedt:07.1}

\begin{figure}[ht!]
\begin{center}
\includegraphics[scale=0.56]{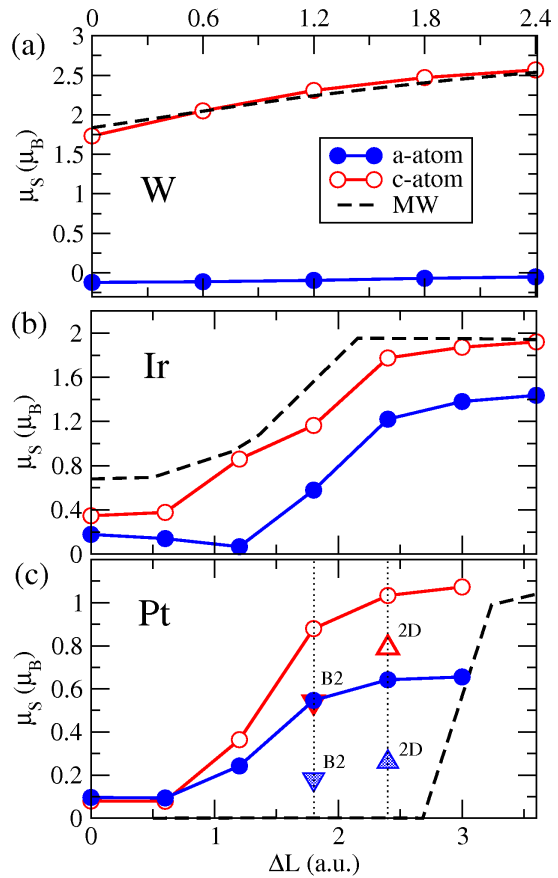}
\end{center}
\caption{\label{fig2} (color online) Local spin moments of
the central ($\mu_S^c$) and apex ($\mu_S^a$) atoms (denoted as
"c-atom" and "a-atom", respectively) of the 
suspended (a) W, (b) Ir and (c) Pt trimers as a function of 
the stretching of the contacts (without SOC). For comparison 
the values for an infinite free monowire (MW) are given by 
dashed lines. The interatomic distance in the infinite MW 
corresponds to the distance $d_c$ for a given $\Delta L$. 
For Pt triangles down and triangles up stand for the values 
of the apex (filled) and central (open) spin moments in B2 
and 2D cases, respectively. For details see text.}
\end{figure}

The magnetic properties of suspended $5d$ TM atoms
are very sensitive to the local environment. In most
previous studies it was common to model the contacts by
semi-infinite surface electrodes.\cite{Smelova:08.1,
Czerner:08.1,Smogunov:08.2} This restriction on the geometry
of the contacts is quite strong in particular for
$5d$ TM break junctions, as the surface geometry leads to
suppressed spin moments of the atoms in suspended wire and
contact atoms closest to it.\cite{Smogunov:08.2}
However, it is well-established that the shape of
the contacts in a break junction is rather a thinning wire-like
geometry, with the reduced coordination number of the contact
atoms adjacent to the chain which greatly enhances their
tendency towards magnetism.\cite{Rego:03.1,Cheng:06.1}
In this respect our contact geometry is probably
closer to the real situation, although it might overestimate
the tendency of the contacts to magnetism.
In order to investigate this point quantitatively, we have
performed calculations in the setup shown in Fig.~2,
which are discussed further below.

The calculated spin moments $\mu_S$ inside the atomic 
spheres of the suspended trimers upon increasing the distance 
between the contacts $\Delta L$ are shown in Fig.~\ref{fig2}. 
For W, the central atom reveals a sizable magnetic moment 
of 1.7$\mu_B$ already at $\Delta L =0$ which further 
increases upon stretching, while the moments of all other 
atoms are negligible. In this respect the W trimer presents 
a unique system of a single spin impurity between nonmagnetic 
leads. At small stretching, the spin moments of the Ir trimer 
are rather small with significantly larger moments of the 
central atom. Upon further stretching the apex and central 
moments rise rapidly, reaching large values of 1.4$\mu_B$ 
and 2$\mu_B$, respectively. Both, the $\mu_S^c$ spin 
moments of the central W and Ir chain atom follow the 
behavior in an infinite monowire very closely as seen in 
Fig.~\ref{fig2}(a) and (b).

The Pt trimer shows a trend similar to Ir upon stretching, 
only with smaller moments $-$ this is in contrast to an 
infinite Pt MW, where the spin moment is zero for a
large interval of interatomic distances without SOC. 
It is resurrected upon including SOC, but only when the
magnetization points along the $z$-axis, an effect
coined as colossal magnetic anisotropy.\cite{Smogunov:08.1}
In contrast, for the suspended Pt trimer, the moments of
the trimer atoms are non-zero already without SOC
for a wide range of $d_c$. Upon including SOC the spin 
moments do not change dramatically and the moments 
are non-zero for both magnetization directions in the 
whole range of $\Delta L$, which manifests the subtle 
magnetism of Pt chains and its sensitivity to local 
environment.

In the critical case of Pt, we analyze in more detail the 
influence of the contact geometry on the spin moments 
of the trimer. At the stretching of $\Delta L=1.8$~a.u.~we 
performed a calculation in which we artificially quenched 
the spin moments in the contacts by applying a small 
magnetic field inside the muffin-tin spheres of the Pt atoms, 
situation referred to as "B2" in the following (see Fig.~2). 
We observe that $\mu_S^a$ drops significantly from 
0.55$\mu_B$ to 0.18$\mu_B$, while $\mu_S^c$ drops 
from 0.89$\mu_B$ to 0.54$\mu_B$ (Fig.~3(c)) $-$ a 
value, very close to that reported 
in~Ref.~[\onlinecite{Smogunov:08.2}], where contacts 
were modelled by infinite bulk electrodes. However, 
quenching the spin moment in the deeper parts of the 
contacts only ("B1"-case, Fig.~2) $-$ the situation which 
is probably closer to that in real experiments $-$ almost 
does not affect the values of $\mu_S^c$ and $\mu_S^a$. 

Finally, we compare the results obtained within our 1D geometry 
of the electrodes with the situation of surface-like electrodes, 
as shown in Fig.~2(a). For a separation of $\Delta L=2.4$~a.u. 
$\mu_S^c$ is reduced in the 2D geometry by only 25\% from 
1.03$\mu_B$ in the wire geometry to 0.79$\mu_B$, while the 
moment of the apex atom is affected stronger, dropping from 
0.64$\mu_B$ to 0.26$\mu_B$ (Fig.~3(c)). In a more realistic 
situation the coordination of the contact atoms next to the 
apex atoms is more reduced and the moments of the trimer's 
atoms will increase approaching the values obtained
in our wire geometry. However, we have to conclude that 
magnetism in suspended short Pt chains with their small 
moments is extremely sensitive to slightest changes in 
the contact geometry. Therefore, observing this magnetism 
experimentally will be a hard task, even considering the 
cluster's large magnetic anisotropy energies.

\begin{figure}
\begin{center}
\includegraphics[scale=0.23]{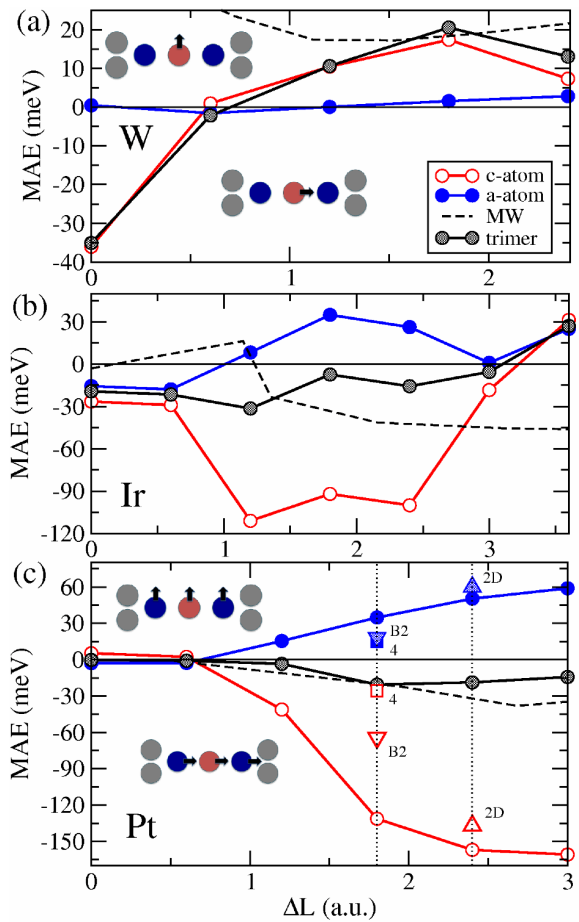}
\end{center}
\caption{\label{fig3} (color online) Magnetic anisotropy
energies of suspended trimers of (a) W, (b) Ir and (c) Pt 
as a function of the stretching of the contacts $\Delta L$.
The total trimer's anisotropy $\Delta E^{\rm tri}$ (black
half-filled circles) is decomposed into the contribution
from the central $\Delta E^{c}$ (red open circles) and
apex atoms $\Delta E^{a}$ (blue filled circles) (for
definitions of the quantities see text). Black dashed
line stands for the MAE of the infinite MWs (per atom).
For Pt squares stand for the $\Delta E^{a}$ (filled)
and $\Delta E^{c}$ (open) tetramer values, while 
triangles up and triangles down stand for the 
corresponding values in B2 and 2D cases. 
For details see text.}
\end{figure}

\section{Magnetic Anisotropy Energies}

Recent theoretical studies,\cite{Thiess:08.1,Silva} which 
are in accordance with experiments,\cite{Kizuka,Legolas} 
clearly state that the interatomic distance $d_c$ in suspended 
chains of $5d$ elements consisting of several atoms can 
reach large values of 5.0$-$5.5~a.u.~upon stretching, which 
corresponds to $\Delta L\approx 2-3$~a.u.,~c.f.~Fig.~\ref{fig1}.
Therefore, in a real Pt, Ir, or W break junction experiment 
the central atom will reveal strong fingerprints of 
spin-polarization at sufficiently small temperatures. In order 
to study the thermal stability of magnetism in the suspended 
trimers, next we concentrate on their magnetic anisotropy 
energies.

In a transport break junction experiment, the data is obtained 
by averaging over thousands of measurements, which differ 
from each other by the thinning history of the wire and the 
contact geometry. Therefore, in order to analyze this data 
with respect to the magnetism of the suspended
chain, it is necessary to disentangle the contributions of
intrinsic trimer's magnetic anisotropy energy $\Delta E^{\rm tri}$ 
from that originating from the contacts. In our calculations, we 
do this by switching SOC off in the contact's atomic spheres.
Moreover, we individually switch SOC on and off in the apex
and central atoms, to determine their separate contributions,
$\Delta E^a$ and $\Delta E^c$, respectively, to the total MAE 
of the trimer, $\Delta E^{\rm tri}$. For W with essentially one 
magnetic atom in the trimer we define $\Delta E^{\rm tri}$ as 
$\Delta E^{c}+2\Delta E^{a}$, while for Ir and Pt we define 
the total MAE of the trimer per atom as 
$\Delta E^{\rm tri}=(\Delta E^{c} +2\Delta E^{a})/3$.
\cite{footnote_mae}

The absolute value of trimer's MAE, shown in Fig.~4 with a black 
solid line for W, Ir and Pt, reach large absolute values of 10 to 30 meV 
per atom at most electrode separations. These values, which are one to
two orders of magnitude larger than those in most of transition-metal 
nanostructures, give us confidence that the magnetism of suspended 
trimers can be tackled experimentally. For all three elements the 
calculated values of anisotropy energies $\Delta E^{\rm tri}$ are 
of the order of magnitude of those predicted for infinite monowires 
(dashed line in Fig.~4 and Refs.~[\onlinecite{Mokrousov:06.3,
Smogunov:08.1}]). Interestingly, the trimer's MAE displays a 
non-trivial dependence on $\Delta L$ quite different from that 
in ideal chains for Ir and W, while for Pt both anisotropy energies 
are close in their trend.

For W, Fig.~4(a), the only contribution to $\Delta E^{\rm tri}$ stems
from the strongly spin-polarized central atom. Starting
at about 35~meV and a magnetization along the chain, there 
is a switch to a perpendicular direction upon stretching
by $\Delta L = 0.6$~a.u. Upon further increasing the
contact separation, the perpendicular magnetization is stabilized
by a sizable MAE of 10$-$20~meV. A switch of the magnetization
direction can be also observed for Ir at significant
stretching of around $\Delta L = 3.0$~a.u., Fig.~4(b). In general,
for Ir, the behavior of $\Delta E^{\rm tri}$ upon stretching
is quite smooth, and the $z$-direction of the magnetization
is favored by considerable 15$-$30~meV over a wide interval
of contact separations.

Detecting traces of magnetism in Pt break junctions at small
stretching will present a significant challenge as can be
seen from Fig.~4(c). In the regime of $\Delta L < 1.3$~a.u.~the
value of $\Delta E^{\rm tri}$ is on the order of a few meVs,
posing the question of whether the sensitive magnetization
will survive temperature fluctuations. Moreover, in this
regime, as pointed out above, the spin moments of the apex 
and central atoms depend strongly on contact geometry and 
details of the thinning process. This will influence the
contribution of these atoms to $\Delta E^{\rm tri}$, causing
frequent measurement-to-measurement oscillations in its
sign and magnitude. Beyond a distance of $\Delta L=1.5$~a.u.
the well-developed magnetization of the trimer
is pointing rigidly along the
chain axis with a MAE of about 15~meV $-$ a value somewhat
smaller than that of an infinite MW of corresponding
interatomic distance, c.f.~Fig.~4(c). 

As a general trend, we observe in Fig.~4 competing
contributions to the MAE from central and apex atoms for
Ir and Pt atomic junctions. At most separations a
positive value from the apex atom, $\Delta E^{a}>0$,
favors a perpendicular magnetization direction, while negative
contribution from the central atom, $\Delta E^{c}<0$,
forces the magnetization along the trimer's axis. Due
to this competition the trimer's MAE behaves
qualitatively differently from that in the infinite MW
as a function of the interatomic distance for Ir.
It significantly quenches the central atom's MAE, 
so that the resulting total anisotropy is on average smaller 
than anticipated from the infinite MW both for Ir and Pt.

The origin of this competition is the different local
symmetry of apex and central atoms and it exists also
in longer chains or chains made of other elements.
To demonstrate this point we additionally calculated the
MAE of a suspended four-atom chain of Pt atoms with the
contacts as in Fig.~1. The $\Delta E^a$ and $\Delta E^c$
anisotropy energies for the two apex and two central atoms,
respectively, are displayed by squares in Fig.~4(c)~(per atom) 
and reveal the same effect: $\Delta E^a$ and $\Delta E^c$ 
are close in their values but opposite in sign.
Due to its large magnitude, this effect may even compete 
with exchange interactions in suspended small chains and 
lead to a non-collinear magnetic order. With increasing
chain length the anisotropy energies of the atoms in the
center of the chain will eventually approach that of  
an atom in the infinite monowire (this can be already
seen in Fig.~4(c) for Pt when going from a trimer to a 
tetramer), and the whole cluster will behave as a
superparamagnet. For smaller suspended clusters, however, 
which are much more probable to occur in an experiment,
the effect of competition between central and apex atoms 
will be dramatic. 

\begin{figure}[htb!]
\begin{center}
\vspace{0.1cm}\includegraphics[scale=0.24]{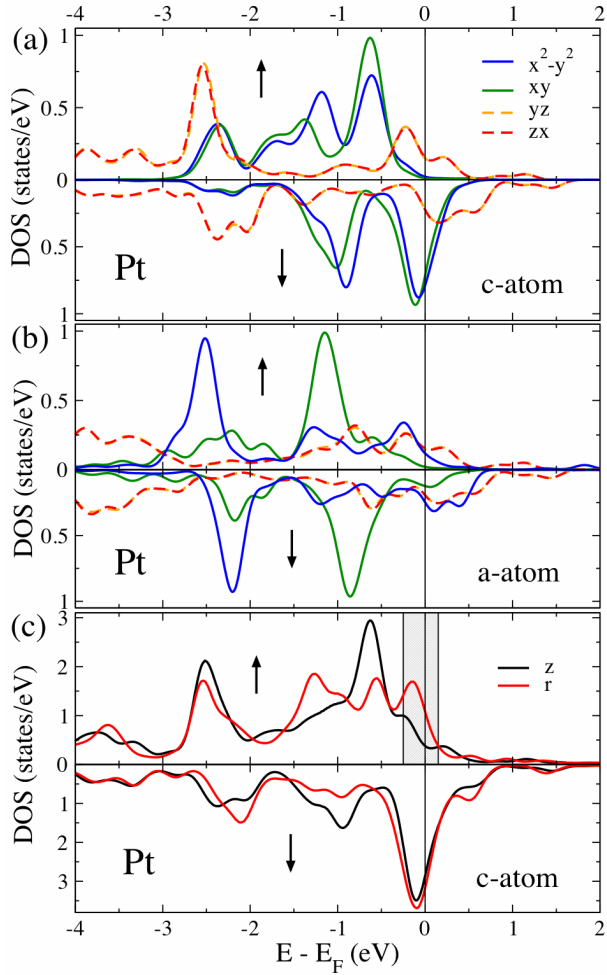}
\end{center}
\caption{\label{fig4} (color online) Spin-decomposed local
density of states (LDOS) for a Pt trimer including SOC for a 
separation between the electrodes of $\Delta L=1.8$~a.u.~(a) 
and (b): $\Delta_3$ ($d_{xz}$,$d_{yz}$) and
$\Delta_4$ ($d_{x^2-y^2}$,$d_{xy}$) $3d$-LDOS of the
central ("c-atom") and apex ("a-atom") atoms for the 
magnetization along the $z$-axis. (c) LDOS of the central 
atom for two different magnetization directions ($z$ and $r$).}
\end{figure}

We demonstrate that the effect of the competition between
the apex and central atoms for the MAE value is also stable 
with respect to the geometry of the contacts. For this purpose 
we calculate the MAE of the Pt trimer in 2D-configuration, as 
well as in B1 and B2 cases, and present the calculated values 
in Fig.~4(c). As far as the 2D-case is concerned, we observe 
that despite significant changes in the spin moments,~c.f.~Fig.~3(c), 
the values of $\Delta E^a$ and $\Delta E^c$ are very close
to those calculated within the wire geometry, moreover, they 
are also opposite in sign. Their competition results in rather 
close values for the total $\Delta E^{\rm tri}$ of 7~and~18~meV 
for the 2D and wire-like geometry of the contacts, respectively, 
both favoring the $z$-direction of the trimer's magnetization.
Notably, both values are very close to that of 12~meV~reported by 
Smogunov~{\it et al.} in Ref.~[\onlinecite{Smogunov:08.2}] for
the suspended between semi-infinite electrodes Pt trimer,
although one has to keep in mind the difference between 
the geometries of all three cases. 

In the B1-situation, in which we do not quench the spin moments 
of the contact atoms of the electrode, the MAE is almost not 
affected compared to the wire setup of Fig.~1. On the other hand, 
in the B2-case, in which we quench the spin moments inside the 
entire electrode, the spin moments of the trimer atoms are strongly 
reduced, c.f.~Fig.~3(c), and so is the MAE: $\Delta E^c$ and 
$\Delta E^a$ drop to almost half their value. However, the latter 
two contributions to the total MAE $\Delta E^{\rm tri}$ are still of 
opposite sign, proving the generality of the phenomenon of 
competing anisotropies for this type of atomic junctions. The 
latter effect is responsible for the value of the total MAE 
$\Delta E^{\rm tri}$ of 9~meV, smaller than the value of 
20~meV obtained within the wire-geometry.

To understand the effect, we analyze the local density
of states (LDOS) and choose the Pt trimer as an example.
In an infinite MW, the position of the localized
$d$-states with respect to the Fermi energy ($E_F$) is
responsible for the formation of the orbital moment and
the direction of the
magnetization.\cite{Mokrousov:06.3,vdLaan:98.1,
Thiess:09.1} According to the symmetry these $d$-states
can be subdivided into $\Delta_3$ ($d_{xz}$,$d_{yz}$) 
and $\Delta_4$ ($d_{x^2-y^2}$,$d_{xy}$) groups,
not taking into account $\Delta_1$ $sd$-hybrid.\cite{Mokrousov:06.3}
In a Pt (Ir) infinite MW, the position of the X-edge ($\Gamma$-edge)
of the $\Delta_4$ degenerate $d_{xy}$ and $d_{x^2-y^2}$
bands at $E_F$ is responsible for the easy
magnetization direction along the chain.\cite{vdLaan:98.1,
Thiess:09.1} In a trimer at larger stretching, the
$5d$-LDOS of a central atom qualitatively follows that
of an atom in an infinite monowire (not shown), which
explains the same sign of the MAE (Fig.~\ref{fig3}).
A small splitting between the $d_{x^2-y^2}$ and $d_{xy}$ states 
due to the presence of the leads can be already seen for the 
central atom (Fig.~5(a)), however, for the apex atom the effect is 
dramatic and has crucial consequences for the apex MAE.
Interaction with the contacts locally breaks the $C_{1\infty}$ 
symmetry of an infinite chain which results in large splitting 
between the $d_{xy}$- and $d_{x^2-y^2}$-orbitals in the
$\Delta_4$-band and their shift towards the lower energies
(Fig.~5(b)). On the other hand, the degeneracy between the
$d_{xz}$ and $d_{yz}$-states ($\Delta_3$-band) is locked by 
symmetry. This leads to the rearrangement of the states 
around $E_F$ and as a result the in-plane direction of the 
apex spin moment becomes favorable (Fig.~4).

In general, it should be possible to deduce the predicted 
switches in the magnetization direction of the suspended chains 
from transport measurements in these systems. As an example,
in Fig.~5(c) we present the LDOS of the central atom in
a Pt trimer at $\Delta L=1.8$~a.u. Upon changing the
magnetization direction significant changes in the LDOS
around the Fermi energy $E_F$ can be observed (shaded
area in Fig.~5(c)), which will inevitably result in
large changes of the experimentally measured conductance.
As far as the $5d$ TMs are concerned, giant values of
the ballistic anisotropic magnetoresistance should be
achievable due to strong spin-orbit coupling in these
metals, which is able to modify the electronic structure
significantly in response to the changes of the
magnetization direction in real space. 

\section{Conclusions}

By performing {\it ab initio} calculations of suspended
trimers of W, Ir and Pt including structural relaxations
as a function of the separation between the leads, we
demonstrate that the chain atoms develop significant 
spin moments upon stretching. We investigate the stability
of these spin moments via evaluating the
magnetic anisotropy energy of the trimers. Our calculations
reveal large MAE of the whole trimers of the order of
10 to 30 meV per atom. Interestingly, we observe that
the total MAE presents a competition between large 
contributions from the apex and central atoms.
We argue that this effect is general and can occur
in suspended chains of different elements and different
length, leading to non-trivial real space textures of magnetic
anisotropy energy which might even lead to complex 
magnetic ordering in atomic-scale junctions.

Under the condition of large predicted values of MAE
huge magnetic fields would be required to 
change the magnetization of the chains, therefore,
new ways to achieve this goal have to be
established. One of the possibilities lies in using
the ability to control the magnetization direction
in the junction by changing the distance between
the leads. We show that such switches 
of the magnetization can happen and will result in 
strong features in the measured conductance. 
Distinguishing these features from the changes in 
conductance due to chain elongation or atomic 
rearrangements presents a considerable challenge.

Financial support of the the Stifterverband f\"ur die 
Deutsche Wissenschaft is gratefully acknowledged. We also thank
Phivos Mavropoulos for discussions.


\begin{thebibliography}{39}
\bibitem{footnote_mae} In reality the value of the apex MAE is obtained by 
     substracting the MAE of the central atom from the MAE of the whole trimer.
\end{thebibliography}
\end{document}